\documentstyle[onecolumn]{mn}

\newif\ifAMStwofonts

\ifoldfss
  \ifCUPmtlplainloaded \else
    \NewTextAlphabet{textbfit} {cmbxti10} {}
    \NewTextAlphabet{textbfss} {cmssbx10} {}
    \NewMathAlphabet{mathbfit} {cmbxti10} {} 
    \NewMathAlphabet{mathbfss} {cmssbx10} {} 
  \fi
  \ifAMStwofonts
    \ifCUPmtlplainloaded \else
      \NewSymbolFont{upmath} {eurm10}
      \NewSymbolFont{AMSa} {msam10}
      \NewMathSymbol{\upi}     {0}{upmath}{19}
      \NewMathSymbol{\umu}     {0}{upmath}{16}
      \NewMathSymbol{\upartial}{0}{upmath}{40}
      \NewMathSymbol{\leqslant}{3}{AMSa}{36}
      \NewMathSymbol{\geqslant}{3}{AMSa}{3E}

      \let\leq=\leqslant 
      \let\geq=\geqslant 
    \fi
  \fi
\fi 

\ifnfssone
  \newmathalphabet{\mathit}
  \addtoversion{normal}{\mathit}{cmr}{m}{it}
  \addtoversion{bold}{\mathit}{cmr}{bx}{it}
  \newmathalphabet{\mathbfit} 
  \addtoversion{normal}{\mathbfit}{cmr}{bx}{it}
  \addtoversion{bold}{\mathbfit}{cmr}{bx}{it}
  \newmathalphabet{\mathbfss} 
  \addtoversion{normal}{\mathbfss}{cmss}{bx}{n}
  \addtoversion{bold}{\mathbfss}{cmss}{bx}{n}
  \ifAMStwofonts
    \ifCUPmtlplainloaded \else
      \UseAMStwoboldmath
      \makeatletter
      \new@mathgroup\upmath@group
      \define@mathgroup\mv@normal\upmath@group{eur}{m}{n}
      \define@mathgroup\mv@bold\upmath@group{eur}{b}{n}
      \edef\UPM{\hexnumber\upmath@group}
      \new@mathgroup\amsa@group
      \define@mathgroup\mv@normal\amsa@group{msa}{m}{n}
      \define@mathgroup\mv@bold\amsa@group{msa}{m}{n}
      \edef\AMSa{\hexnumber\amsa@group}
      \makeatother
      \mathchardef\upi="0\UPM19
      \mathchardef\umu="0\UPM16
      \mathchardef\upartial="0\UPM40
      \mathchardef\leqslant="3\AMSa36
      \mathchardef\geqslant="3\AMSa3E

      \let\leq=\leqslant 
      \let\geq=\geqslant 
    \fi
  \fi
\fi 

\ifnfsstwo
  \DeclareMathAlphabet{\mathbfit}{OT1}{cmr}{bx}{it}
  \SetMathAlphabet\mathbfit{bold}{OT1}{cmr}{bx}{it}
  \DeclareMathAlphabet{\mathbfss}{OT1}{cmss}{bx}{n}
  \SetMathAlphabet\mathbfss{bold}{OT1}{cmss}{bx}{n}
  \ifAMStwofonts
    \ifCUPmtlplainloaded \else
      \DeclareSymbolFont{UPM}{U}{eur}{m}{n}
      \SetSymbolFont{UPM}{bold}{U}{eur}{b}{n}
      \DeclareSymbolFont{AMSa}{U}{msa}{m}{n}
      \DeclareMathSymbol{\upi}{0}{UPM}{"19}
      \DeclareMathSymbol{\umu}{0}{UPM}{"16}
      \DeclareMathSymbol{\upartial}{0}{UPM}{"40}
      \DeclareMathSymbol{\leqslant}{3}{AMSa}{"36}
      \DeclareMathSymbol{\geqslant}{3}{AMSa}{"3E}

      \let\leq=\leqslant 
      \let\geq=\geqslant 
    \fi
  \fi
\fi 

\ifCUPmtlplainloaded \else
  \ifAMStwofonts \else 
    \def\upi{\pi}
    \def\umu{\mu}
    \def\upartial{\partial}
  \fi
\fi

\title[Relativistic corrections to the Sunyaev-Zel'dovich effect]
{Relativistic corrections to the multiple scattering effect on the Sunyaev-Zel'dovich effect in the isotropic approximation}
\author[N. Itoh et al.]
  {Naoki Itoh,$^1$\thanks{Email:n$\_$itoh@hoffman.cc.sophia.ac.jp} 
  Youhei Kawana,$^1$\thanks{Email:y-kawana@hoffman.cc.sophia.ac.jp} Satoshi Nozawa,$^2$\thanks{Email:snozawa@josai.ac.jp} and Yasuharu Kohyama,$^3$\thanks{Email:kohyama@star.fuji-ric.co.jp} \\
  $^{1}$Department of Physics, Sophia University, 7-1 Kioi-cho, Chiyoda-ku, Tokyo, 102-8554, Japan\\
  $^{2}$Josai Junior College for Women, 1-1 Keyakidai, Sakado-shi, Saitama, 350-0295, Japan\\
  $^{3}$Fuji Research Institute Corporation, 2-3 Kanda-Nishiki-cho, Chiyoda-ku, Tokyo, 101-8443, Japan}
\date{Submitted 2001 February}
\pagerange{\pageref{firstpage}--\pageref{lastpage}}
\pubyear{2001}

\def\LaTeX{L\kern-.36em\raise.3ex\hbox{a}\kern-.15em
    T\kern-.1667em\lower.7ex\hbox{E}\kern-.125emX}

\begin{document}

\label{firstpage}

\maketitle

\begin{abstract}
  We extend the formalism for the calculation of the relativistic corrections to the Sunyaev-Zel'dovich effect for clusters of galaxies and include the multiple scattering effects in the isotropic approximation.  We present the results of the calculations by the Fokker-Planck expansion method as well as by the direct numerical integration of the collision term of the Boltzmann equation.  The multiple scattering contribution is found to be very small compared with the single scattering contribution.  For high-temperature galaxy clusters of $k_{B} T_{e} \approx 15$keV, the ratio of the both contributions is $-0.2\%$ in the Wien region.  In the Rayleigh--Jeans region the ratio is  $-0.03\%$.  Therefore the multiple scattering contribution is safely neglected for the observed galaxy clusters.
\end{abstract}

\begin{keywords}
cosmic microwave background --- cosmology: theory --- galaxies: clusters: general --- radiation mechanisms: thermal --- relativity.
\end{keywords}

\section{Introduction}

  Compton scattering of the cosmic microwave background (CMB) radiation by hot intracluster gas --- the Sunyaev-Zel'dovich effect (Zel'dovich \& Sunyaev 1969; Sunyaev \& Zel'dovich 1972, 1980a, 1980b, 1981) --- provides a useful method to measure the Hubble constant $H_{0}$ (Gunn 1978; Silk \& White 1978; Birkinshaw 1979; Cavaliere, Danese \& De Zotti 1979; Birkinshaw, Hughes \& Arnaud 1991; Birkinshaw \& Hughes 1994; Myers et al. 1995; Herbig et al. 1995; Jones 1995; Markevitch et al. 1996; Holzapfel et al. 1997; Hughes \& Birkinshaw 1998; Furuzawa et al. 1998; Komatsu et al. 1999; Reese et al. 2000).  The original Sunyaev-Zel'dovich formula has been derived from a kinetic equation for the photon distribution function taking into account the Compton scattering by electrons: the Kompaneets equation (Kompaneets 1957; Weymann 1965).  The original Kompaneets equation has been derived with a nonrelativistic approximation for the electron.  However, recent X-ray observations have revealed the existence of many high-temperature galaxy clusters (David et al. 1993; Arnaud et al. 1994; Markevitch et al. 1994; Mushotzky \& Scharf 1997; Markevitch 1998).  In particular, Tucker et al. (1998) reported the discovery of a galaxy cluster with the electron temperature $k_{B} T_{e} = 17.4 \pm 2.5$ keV.  Rephaeli and his collaborator (Rephaeli 1995; Rephaeli \& Yankovitch 1997) have emphasized the need to take into account the relativistic corrections to the Sunyaev-Zel'dovich effect for clusters of galaxies.

  In recent years remarkable progress has been achieved in the theoretical studies of the relativistic corrections to the Sunyaev-Zel'dovich effects for clusters of galaxies.  Stebbins (1997) generalized the Kompaneets equation.  Itoh, Kohyama \& Nozawa (1998) have adopted a relativistically covariant formalism to describe the Compton scattering process (Berestetskii, Lifshitz \& Pitaevskii 1982; Buchler \& Yueh 1976), thereby obtaining higher-order relativistic corrections to the thermal Sunyaev-Zel'dovich effect in the form of the Fokker-Planck expansion.  In their derivation, the scheme to conserve the photon number at every stage of the expansion which has been proposed by Challinor \& Lasenby (1998) played an essential role.  The results of Challinor \& Lasenby (1998) are in agreement with those of Itoh et al. (1998).  The latter results include higher-order expansions.  Itoh et al. (1998) have also calculated the collision integral of the Boltzmann equation numerically and have compared the results with those obtained by the Fokker-Planck expansion method.  They have confirmed that the Fokker-Planck expansion method gives an excellent result for $k_{B}T_{e} \leq 15$keV, where $T_{e}$ is the electron temperature.  For $k_{B}T_{e} \geq 15$keV, however, the Fokker-Planck expansion results show nonnegligible deviations from the results obtained by the numerical integration of the collision term of the Boltzmann equation.

  In our previous papers devoted to the study of the relativisitc corrections to the Sunyaev-Zel'dovich effect for clusters of galaxies (Itoh, Kohyama, \& Nozawa 1998; Nozawa, Itoh, \& Kohyama 1998; Itoh, Nozawa, \& Kohyama 2000; Nozawa et al. 2000), we have so far restricted ourselves to the case of single Compton scattering.  This is justified because the optical depth for the Compton scattering of the CMB photon inside the galaxy clusters is generally about $10^{-2}$ or smaller (Birkinshaw 1999).  Nevertheless, it would be desirable to evaluate the effects of the multiple Compton scattering of the CMB photon inside the galaxy clusters accurately, as we have already developed the method to calculate the relativistic corrections to the Sunyaev-Zel'dovich effect for the galaxy clusters with high accuracy.  The multiple scattering effects have been already considered by many authors (Wright 1979; Fabbri 1981; Loeb, McKee \& Lahav 1991; Sazonov \& Sunyaev 1998; Molnar \& Birkinshaw 1999; see Birkinshaw 1999 for other references).  Molnar \& Birkinshaw (1999), in particular, have carried out a detailed Monte Carlo calculation including multiple scattering effects.  However, most of the calculations to date other than Monte Carlo type have assumed isotropy of the radiation field after the first Compton scattering.  In this paper we wish to evaluate the multiple scattering effects in the same theoretical framework of our previous papers.  As a matter of fact, the method of calculating the multiple scattering contributions to the Sunyaev-Zel'dovich effect adopted by Fabbri (1981) and also by Sazonov \& Sunyaev (1998) is in the same line as the present paper.  The lowest-order term has been already obtained by them.  We will calculate the relativistic corrections (higher-order terms) in the present paper.  We will also carry out direct numerical integration of the collision term of the Boltzmann equation and compare the results with those obtained by the Fokker-Plank expansion method.  The present paper is complementary to the work of Molnar \& Birkinshaw (1999) who presented Monte Carlo results without the isotropic approximation.  In the central region of a spherical galaxy cluster, the assumption of the isotropy of the incident radiation field is valid.  Therefore, rigorously speaking, the present result is valid for such conditions of restricted geometry.  Nevertherless, the analytical as well as numerical results with the isotropic approximation in the present paper will be useful when one compares with the full numerical calculation which does not assume the isotropy of the incident radiation field.

  The present paper is organized as follows.  In $\S$ 2 we give the method of the calculation and the results.  In $\S$ 3 we give discussion of the results and concluding remarks.

\section{Mutiple scattering contribution in the isotropic approximation}

  In the present paper, we would like to derive the analytic as well as numerical expressions for the multiple scattering contribution to the Sunyaev-Zeldovich effect for the central region of a spherical galaxy cluster in the isotropic approximation.  As a reference system, we choose the system that is fixed to the center of mass of the galaxy cluster.  The galaxy cluster is assumed to be fixed to the cosmic microwave background (CMB).  Following Itoh et al. (1998), we start with the Fokker-Planck expansion for the time evolution equation of the CMB photon distribution function $n(\omega)$:
\begin{eqnarray}
\frac{ \partial n(\omega)}{ \partial t} & = & 
 2 \left[ \frac{ \partial n}{ \partial x} + n(1+n) \right] \, I_{1} 
  \nonumber  \\
& + & 2 \left[ \frac{ \partial^{2} n}{ \partial x^{2}} 
+ 2(1+n) \frac{ \partial n}{ \partial x} + n(1+n)  \right] \, I_{2}
  \nonumber  \\
& + & 2 \left[\frac{ \partial^{3} n}{ \partial x^{3}} 
+ 3(1+n) \frac{ \partial^{2} n}{ \partial x^{2}} 
+ 3(1+n) \frac{ \partial n}{ \partial x} + n(1+n)  \right] \, I_{3}
  \nonumber \\
& + & 2 \left[\frac{ \partial^{4} n}{ \partial x^{4}} 
+ 4(1+n) \frac{ \partial^{3} n}{ \partial x^{3}}
+ 6(1+n) \frac{ \partial^{2} n}{ \partial x^{2}}
+ 4(1+n) \frac{ \partial n}{ \partial x} + n(1+n)  \right] \, I_{4}
  \nonumber \\
& + & \cdot \cdot \cdot  \, \, \, ,
\end{eqnarray}
where 
\begin{eqnarray}
x  & \equiv & \frac{\omega}{k_{B} T_{e}}  \, ,  \\
\Delta x  & \equiv & \frac{ \omega^{\prime} - \omega}{k_{B} T_{e}}  \, ,  \\
I_{k} & \equiv &  \frac{1}{k !} \int \frac{d^{3}p}{(2\pi)^{3}} d^{3}p^{\prime} d^{3}k^{\prime} \, W \, f(E) \, (\Delta x)^{k} \, .
\end{eqnarray}
In equation (4), $W$ is the transition probability of the Compton scattering, $f(E)$ is the relativistic Maxwellian distribution function for electrons with temperature $T_{e}$.
We have integrated equation (4) analytically with power series expansions of the integrand.  The expansion parameter is 
\begin{equation}
\theta_{e} \equiv \frac{k_{B} T_{e}}{mc^{2}}  \, .
\end{equation}
The explicit forms for $I_{k}$ are given in Itoh et al. (1998).

  We first assume the initial photon distribution of the CMB radiation to be Planckian with temperature $T_{0}$:
\begin{eqnarray}
n (X) & = & n_{0} (X) \, \equiv \, \frac{1}{e^{X} - 1} \, ,
\end{eqnarray}
where
\begin{eqnarray}
X & \equiv & \frac{\omega}{k_{B} T_{0}}  \, .
\end{eqnarray}
Assuming $T_{0}/T_{e} \ll 1$, one obtains the following expression for the fractional distortion of the photon spectrum derived by Itoh et al. (1998):
\begin{eqnarray}
\frac{\Delta n(X)}{n_{0}(X)} & = & \frac{y \, \theta_{e} X e^{X}}{e^{X}-1} \, \left[  \,
Y_{0} \, + \, \theta_{e} Y_{1} \, + \, \theta_{e}^{2} Y_{2} \, + \,  \theta_{e}^{3} Y_{3} \, + \,  \theta_{e}^{4} Y_{4} \, + \,  \theta_{e}^{5} Y_{5} \, + \,  \theta_{e}^{6} Y_{6} \, \right]  \, ,
\end{eqnarray}
\begin{eqnarray}
y & \equiv & \sigma_{T} \int_{0}^{\ell} d \ell_{1} N_{e}(\ell_{1})  \, ,
\end{eqnarray}
where $\sigma_{T}$ is the Thomson scattering cross section, $N_{e}$ is the electron number density, and the integral is over the photon path length in the cluster.  The explicit forms for $Y_{0}, Y_{1}, Y_{2}, Y_{3}$ and $Y_{4}$ are given in Itoh et al. (1998).  In the present paper we have also included higher order terms of $\theta_{e}^{5} Y_{5}$ and $\theta_{e}^{6} Y_{6}$.  The explicit forms are given as follows:
\begin{eqnarray}
Y_{5} & = & - \frac{45}{8} - \frac{7515}{32} \tilde{X} - \frac{28917}{2} \tilde{X}^{2} + \frac{795429}{8} \tilde{X}^{3} - \frac{2319993}{14} \tilde{X}^{4} + \frac{12667283}{112} \tilde{X}^{5} \, \nonumber  \\
&  &  \, \hspace{0.6cm} - \frac{806524}{21} \tilde{X}^{6} + \frac{21310}{3} \tilde{X}^{7} - \frac{46679}{63} \tilde{X}^{8}  + \frac{10853}{252} \tilde{X}^{9} - \frac{58}{45} \tilde{X}^{10} + \frac{29}{1890} \tilde{X}^{11} \,  \nonumber  \\
& + &  \tilde{S}^2 \left( - \frac{28917}{4} + \frac{795429}{4} \tilde{X} - \frac{25519923}{28} \tilde{X}^{2} + \frac{164674679}{112} \tilde{X}^{3} - \frac{7661978}{7} \tilde{X}^{4}  \right.  \, \nonumber  \\
&  &   \left.  \hspace{0.6cm}  + 426200 \tilde{X}^{5} - \frac{11529713}{126} \tilde{X}^{6} + \frac{2724103}{252} \tilde{X}^{7}  - \frac{29377}{45} \tilde{X}^{8} + \frac{14761}{945} \tilde{X}^{9} \right)   \,  \nonumber  \\
& + &  \tilde{S}^{4} \left( - \frac{2319993}{14} + \frac{215343811}{224} \tilde{X} - \frac{12097860}{7} \tilde{X}^{2} + 1363840 \tilde{X}^{3} \right.  \nonumber  \\
&  &  \left.  \hspace{0.6cm}  - \frac{11296318}{21} \tilde{X}^{4} + \frac{18439247}{168} \tilde{X}^{5} - \frac{494392}{45} \tilde{X}^{6} + \frac{400171}{945} \tilde{X}^{7} \right)  \,  \nonumber  \\
& + &  \tilde{S}^{6} \left( - \frac{3427727}{21} + \frac{1321220}{3} \tilde{X} - \frac{25019944}{63} \tilde{X}^{2} + \frac{9756847}{63} \tilde{X}^{3}  \right.  \nonumber  \\
&  &  \left.  \hspace{0.6cm} - \frac{2407609}{90} \tilde{X}^{4} + \frac{3149197}{1890} \tilde{X}^{5} \right)   \,  \nonumber  \\
& + &  \tilde{S}^{8} \left( - \frac{1447049}{63} + \frac{7499423}{252} \tilde{X} - \frac{513242}{45} \tilde{X}^{2} + \frac{4973819}{3780} \tilde{X}^{3} \right)  \,  \nonumber  \\
& + & \tilde{S}^{10} \left( - \frac{20039}{45} + \frac{158369}{945} \tilde{X} \right)    \, ,   \\
  \nonumber  \\
Y_{6} & = &  \frac{7425}{256} + \frac{128655}{1024} \tilde{X} - \frac{360675}{32} \tilde{X}^{2} + \frac{50853555}{128} \tilde{X}^{3} - \frac{45719721}{32} \tilde{X}^{4} + \frac{458203107}{256} \tilde{X}^{5} \, \nonumber  \\
&  &  \hspace{0.6cm}  - \frac{22251961}{21} \tilde{X}^{6} + \frac{71548297}{210} \tilde{X}^{7} - \frac{26865067}{420} \tilde{X}^{8} + \frac{7313155}{1008} \tilde{X}^{9} - \frac{4492}{9} \tilde{X}^{10}  \,  \nonumber  \\
&  &  \hspace{0.6cm}  + \frac{6361}{315} \tilde{X}^{11} - \frac{296}{675} \tilde{X}^{12} + \frac{37}{9450} \tilde{X}^{13}  \, \nonumber  \\
& + &  \tilde{S}^{2} \left( - \frac{360675}{64} + \frac{50853555}{64} \tilde{X} - \frac{502916931}{64} \tilde{X}^{2} + \frac{5956640391}{256} \tilde{X}^{3} - \frac{422787259}{14} \tilde{X}^{4}   \right.  \,  \nonumber  \\
&  &   \hspace{0.6cm} + \frac{143096594}{7} \tilde{X}^{5} - \frac{6635671549}{840} \tilde{X}^{6} + \frac{1835601905}{1008} \tilde{X}^{7} - \frac{2275198}{9} \tilde{X}^{8} + \frac{6475498}{315} \tilde{X}^{9} \,  \nonumber  \\
&  &  \left. \hspace{0.6cm}  - \frac{201428}{225} \tilde{X}^{10} + \frac{50431}{3150} \tilde{X}^{11}    \right)  \nonumber  \\
& + & \tilde{S}^{4} \left( - \frac{45719721}{32} + \frac{7789452819}{512} \tilde{X} - \frac{333779415}{7} \tilde{X}^{2} + \frac{2289545504}{35} \tilde{X}^{3} - \frac{3250673107}{70} \tilde{X}^{4}   \right. \,  \nonumber  \\
&  &  \left.  \hspace{0.6cm}  + \frac{12425050345}{672} \tilde{X}^{5} - \frac{38289808}{9} \tilde{X}^{6} + \frac{175550878}{315} \tilde{X}^{7} - \frac{344840}{9} \tilde{X}^{8} + \frac{1343507}{1260} \tilde{X}^{9}   \right)  \nonumber  \\
& + &  \tilde{S}^{6} \left( - \frac{378283337}{84} + \frac{2217997207}{105} \tilde{X} - \frac{3599918978}{105} \tilde{X}^{2} + \frac{6574526345}{252} \tilde{X}^{3} - \frac{93232583}{9} \tilde{X}^{4}   \right.  \nonumber  \\
&  &  \left.  \hspace{0.6cm} + \frac{690760073}{315} \tilde{X}^{5} - \frac{31433128}{135} \tilde{X}^{6} + \frac{9151432}{945} \tilde{X}^{7}    \right)  \nonumber  \\
& + &  \tilde{S}^{8} \left( - \frac{832817077}{420} + \frac{5053390105}{1008} \tilde{X} - \frac{39749708}{9} \tilde{X}^{2} + \frac{1090981471}{630} \tilde{X}^{3}   \,  \right.  \nonumber  \\
&  &  \left.  \hspace{0.6cm}  - \frac{13818464}{45} \tilde{X}^{4} + \frac{12565681}{630} \tilde{X}^{5}  \right)   \nonumber  \\
& + &  \tilde{S}^{10} \left( - \frac{1551986}{9} + \frac{69474842}{315} \tilde{X} - \frac{19157638}{225} \tilde{X}^{2} + \frac{63282617}{6300} \tilde{X}^{3}   \right)   \nonumber  \\
& + &  \tilde{S}^{12} \left( - \frac{1616456}{675} + \frac{34394053}{37800} \tilde{X}  \right)  \, , 
\end{eqnarray}
where
\begin{eqnarray}
\tilde{X} & \equiv &  X \, {\rm coth} \left( \frac{X}{2} \right)  \, , \\
\tilde{S} & \equiv & \frac{X}{ \displaystyle{ {\rm sinh} \left( \frac{X}{2} \right)} }   \, .
\end{eqnarray}
Equation (8) is the single scattering contribution, i.e. the first-order term in $y$.  If the cluster of galaxies is optically thin, i.e. $y \ll 1$, the single scattering approximation is a good approximation.  In fact, the approximation is valid for most of the clusters.  However, it is extremely important to calculate the next-order contribution in order to obtain more accurate theoretical prediction for the future observation of the Sunyaev-Zeldovich effect for clusters of galaxies.

  We now calculate the multiple scattering contribution.  Since $y \ll 1$ is realized for most of clusters of galaxies, the second-order contribution is considered to be sufficient.  We now assume that the initial photon distribution has an isotropic first-order perturbation.  Namely,
\begin{eqnarray}
n (X) & = & n_{1} (X) \, \equiv \, n_{0}(X) + \Delta n(X) \, , \nonumber  \\
      &   & \, \, \hspace{1.0cm} \, \, = \, n_{0}(X) \left\{ 1 \, + \, \frac{\Delta n(X)}{n_{0}(X)} \right\}  \, ,
\end{eqnarray}
where the second term in equation (14) is given by equation (8).  Here we have followed Fabbri (1981) and Sazonov \& Sunyaev (1998) in assuming that the  radiation field after the first Compton scattering is isotropic.  To go beyond the isotropic approximation would be a very involved calculation depending on the exact geometry of the galaxy cluster.  In such a calculation simple analytic expressions resulting from the Fokker-Planck expansions would not be possible.  In this paper, therefore, we content ourselves with the isotropic approximation and carry out higher-order Fokker-Planck expansions. Inserting equation (14) into RHS of equation (1), and performing the standard calculation, we obtain the following expression for the fractional distortion of the photon distribution function including the second-order contribution:

\begin{eqnarray}
\frac{\Delta n(X)}{n_{0}(X)} & = & \frac{y \, \theta_{e} X e^{X}}{e^{X}-1} \, \left[  \,
Y_{0} \, + \, \theta_{e} Y_{1} \, + \, \theta_{e}^{2} Y_{2} \, + \,  \theta_{e}^{3} Y_{3} \,  + \, \theta_{e}^{4} Y_{4} \, + \, \theta_{e}^{5} Y_{5} \, + \,  \theta_{e}^{6} Y_{6} \, \right]  \, ,  \nonumber  \\
& + &  \frac{1}{2} \, \frac{y^2 \, \theta_{e}^2 X e^{X}}{e^{X}-1} \, \left[  \,
 Z_{0} \, + \, \theta_{e} Z_{1} \, + \, \theta_{e}^{2} Z_{2} \, + \, \theta_{e}^{3} Z_{3} \, + \, \theta_{e}^{4} Z_{4} \, + \, \theta_{e}^{5} Z_{5} \, + \, \theta_{e}^{6} Z_{6} \, \right]  \, ,   \\
  \nonumber  \\
Z_{0} & = & - 16 + 34 \tilde{X} - 12 \tilde{X}^2 + \tilde{X}^3 
             \, + \, \tilde{S}^2 \left( - 6 + 2 \tilde{X} \right)  \, ,  \\ 
  \nonumber  \\
Z_{1} & = &  - 80 + 590 \tilde{X} - \frac{3492}{5} \tilde{X}^2 + \frac{1271}{5} \tilde{X}^3 - \frac{168}{5} \tilde{X}^4 + \frac{7}{5} \tilde{X}^5   \,  \nonumber  \\ 
      & + &  \tilde{S}^2 \left( -\frac{1746}{5} + \frac{2542}{5} \tilde{X} 
                 - \frac{924}{5} \tilde{X}^2 + \frac{91}{5} \tilde{X}^3 \right) \, \nonumber \\
      & + &  \tilde{S}^4 \left( -\frac{168}{5} + \frac{119}{10} \tilde{X} \right)  \,  ,  \\ 
  \nonumber  \\
Z_{2} & = & - 160 + 4792 \tilde{X} - \frac{357144}{25} \tilde{X}^2 
               + \frac{312912}{25} \tilde{X}^3 - \frac{110196}{25} \tilde{X}^4 \, \nonumber \\
      &   & +    \frac{34873}{50} \tilde{X}^5 - \frac{734}{15} \tilde{X}^6 
               + \frac{367}{300} \tilde{X}^7    \,  \nonumber  \\
      & + &   \tilde{S}^2 \left( -\frac{178572}{25} + \frac{625824}{25} \tilde{X} 
              - \frac{606078}{25} \tilde{X}^2 + \frac{453349}{50} \tilde{X}^3 \right.   \,                     \nonumber \\
      &   &   \left. \, \hspace{0.6cm} \, - \frac{20919}{15} \tilde{X}^4 + \frac{367}{5}                \tilde{X}^5 \right)  \,      \nonumber \\       
      & + &   \tilde{S}^4 \left( -\frac{110196}{25} + \frac{592841}{100} \tilde{X} 
                 - 2202 \tilde{X}^2 + \frac{5872}{25} \tilde{X}^3 \right)  \, \nonumber  \\
      & + &   \tilde{S}^6 \left( -\frac{6239}{30} + \frac{11377}{150} \tilde{X} \right) \, ,  \\
  \nonumber  \\
Z_{3} & = & - 90 + \frac{96651}{4} \tilde{X} - \frac{8659449}{50} \tilde{X}^2
                 + \frac{62384943}{200} \tilde{X}^3  - \frac{38586081}{175} \tilde{X}^4  \,      \nonumber  \\
      &   &  +  \frac{103117227}{1400} \tilde{X}^5 - \frac{1325008}{105} \tilde{X}^6  
                 + \frac{590831}{525} \tilde{X}^7 - \frac{1718}{35} \tilde{X}^8
                 + \frac{859}{1050} \tilde{X}^9  \, \nonumber  \\
      & + &  \tilde{S}^2 \left( - \frac{8659449}{100} + \frac{62384943}{100} \tilde{X}
                 - \frac{424446891}{350} \tilde{X}^2 + \frac{1340523951}{1400} \tilde{X}^3   \right.  \, \nonumber  \\
      &   &       \left. \, \hspace{0.6cm} \,  - \frac{12587576}{35} \tilde{X}^4  
              + \frac{2363324}{35} \tilde{X}^5 - \frac{212173}{35} \tilde{X}^6
                +  \frac{215609}{1050} \tilde{X}^7  \right)  \,      \nonumber  \\
      & + &  \tilde{S}^4 \left( - \frac{38586081}{175} + \frac{1752992859}{2800} \tilde{X}  - \frac{3975024}{7} \tilde{X}^2 + \frac{37813184}{175} \tilde{X}^3  \right. \,              \nonumber  \\
      &   &    \left. \, \hspace{0.6cm} \, - \frac{1247268}{35} \tilde{X}^4
               + \frac{1459441}{700} \tilde{X}^5 \right) \,              \nonumber  \\
      & + &  \tilde{S}^6 \left( - \frac{5631284}{105} + \frac{36631522}{525} \tilde{X}
                 - \frac{920848}{35} \tilde{X}^2 + \frac{1544482}{525} \tilde{X}^3  \right) \, ,             \nonumber  \\
      & + &  \tilde{S}^8 \left( - \frac{53258}{35} + \frac{593569}{1050} \tilde{X}  \right) \, ,  \\
  \nonumber  \\
  Z_{4} & = &  60 + 82497 \tilde{X} - \frac{36883086}{25} \tilde{X}^2 
                  + \frac{129233103}{25} \tilde{X}^3
                  - \frac{1154992263}{175} \tilde{X}^4   \, \nonumber  \\
      &   &  +  \frac{5504779501}{1400} \tilde{X}^5
                  - \frac{129898756}{105} \tilde{X}^6 + \frac{114929504}{525} \tilde{X}^7 - \frac{2337809}{105} \tilde{X}^8  \,  \nonumber  \\
      &   &  + \frac{2681837}{2100} \tilde{X}^9
                  - \frac{2851}{75} \tilde{X}^{10} + \frac{2851}{6300} \tilde{X}^{11}  \,       \nonumber  \\
      & + & \tilde{S}^{2} \left( - \frac{18441543}{25} + \frac{258466206}{25} \tilde{X}
                  - \frac{12704914893}{350} \tilde{X}^2 + \frac{71562133513}{1400} \tilde{X}^3   \right.  \, \nonumber  \\
      &   &   \, \hspace{0.6cm} \,  - \frac{1234038182}{35} \tilde{X}^4 + \frac{459718016}{35} \tilde{X}^5 - \frac{577438823}{210} \tilde{X}^6
 + \frac{673141087}{2100} \tilde{X}^7  \,  \nonumber  \\
      &   &   \left. \,  \hspace{0.6cm} \,  - \frac{2888063}{150} \tilde{X}^8  + \frac{1451159}{3150} \tilde{X}^9  \right) \, \nonumber  \\
      & + &  \tilde{S}^4 \left( - \frac{1154992263}{175} + \frac{93581251517}{2800} \tilde{X} - \frac{389696268}{7} \tilde{X}^2 + \frac{7355488256}{175} \tilde{X}^3  \right.           \,  \nonumber  \\
      &   &  \left. \, \hspace{0.6cm} \, - \frac{565749778}{35} \tilde{X}^4 + \frac{4556441063}{1400} \tilde{X}^5  - \frac{24301924}{75} \tilde{X}^6
               + \frac{39340949}{3150} \tilde{X}^7 \right)  \,  \nonumber  \\
      & + &  \tilde{S}^6 \left( - \frac{552069713}{105} + \frac{7125629248}{525} \tilde{X} - \frac{1253065624}{105} \tilde{X}^2 + \frac{2410971463}{525} \tilde{X}^3 \right.        \, \nonumber  \\
      &   &   \left. \, \hspace{0.6cm} \, - \frac{236692871}{300} \tilde{X}^4 + \frac{309598643}{6300} \tilde{X}^5  \right)    \,  \nonumber  \\
      & + &   \tilde{S}^8 \left( - \frac{72472079}{105} + \frac{1853149367}{2100} \tilde{X} - \frac{25228499}{75} \tilde{X}^2 + \frac{488977861}{12600} \tilde{X}^3  \right)        \,  \nonumber   \\
      & + & \tilde{S}^{10} \left( - \frac{1970041}{150} + \frac{15569311}{3150} \tilde{X} \right)     \, ,   \\
  \nonumber  \\
Z_{5} & = & - \frac{135}{8} + \frac{12368565}{64} \tilde{X} 
     - \frac{1523246139}{160} \tilde{X}^{2}
     + \frac{41024053941}{640} \tilde{X}^{3}
     - \frac{78913341669}{560} \tilde{X}^{4}   \nonumber  \\
    &  & + \frac{124274226315}{896} \tilde{X}^{5} 
     - \frac{2505515368}{35} \tilde{X}^{6}
     + \frac{29643451897}{1400} \tilde{X}^{7}
     - \frac{105635617}{28} \tilde{X}^{8}      \nonumber  \\
    &  & + \frac{10436409287}{25200} \tilde{X}^{9}
     - \frac{6284921}{225} \tilde{X}^{10}
     + \frac{2347649}{2100} \tilde{X}^{11}
     - \frac{16312}{675} \tilde{X}^{12}
     + \frac{2039}{9450} \tilde{X}^{13}       \nonumber  \\
   & + & \tilde{S}^{2} \left( - \frac{1523246139}{320} 
            + \frac{41024053941}{320} \tilde{X}
            - \frac{868046758359}{1120} \tilde{X}^{2}
            + \frac{1615564942095}{896} \tilde{X}^{3}  \right.  \nonumber  \\
        &  &  \, \hspace{0.6cm} \,   - \frac{71407187988}{35} \tilde{X}^{4}
            + \frac{88930355691}{70} \tilde{X}^{5}
            - \frac{26091997399}{56} \tilde{X}^{6}
            + \frac{2619538731037}{25200} \tilde{X}^{7}  \nonumber  \\
        &  &  \left. \, \hspace{0.6cm} \,  - \frac{6366624973}{450} \tilde{X}^{8}
            + \frac{1194953341}{1050} \tilde{X}^{9}
            - \frac{11100316}{225} \tilde{X}^{10}
            + \frac{2779157}{3150} \tilde{X}^{11}     \right)   \nonumber  \\
   & + & \tilde{S}^{4} \left( - \frac{78913341669}{560}
               + \frac{2112661847355}{1792} \tilde{X}
               - \frac{22549638312}{7} \tilde{X}^{2}
               + \frac{711442845528}{175} \tilde{X}^{3}  \right. \nonumber  \\
        &  &   \, \hspace{0.6cm} \,  - \frac{38345728971}{14} \tilde{X}^{4}
               + \frac{17731459378613}{16800} \tilde{X}^{5}
               - \frac{53572666604}{225} \tilde{X}^{6}
               + \frac{32395208551}{1050} \tilde{X}^{7}  \nonumber  \\
        &  &  \left.  \, \hspace{0.6cm} \, - \frac{19003480}{9} \tilde{X}^{8}
               + \frac{74038129}{1260} \tilde{X}^{9}   \right)  \nonumber \\
    & + & \tilde{S}^{6} \left( - \frac{10648440314}{35} 
               + \frac{918947008807}{700} \tilde{X}
               - \frac{14155172678}{7} \tilde{X}^{2}
               + \frac{9382331949013}{6300} \tilde{X}^{3}  \right. \nonumber \\
    &  &   \left.  \, \hspace{0.6cm} \, - \frac{521780426341}{900} \tilde{X}^{4}
               + \frac{254938247857}{2100} \tilde{X}^{5}
               - \frac{1732220216}{135} \tilde{X}^{6}
               + \frac{504318104}{945} \tilde{X}^{7}  \right)  \nonumber \\
    & + & \tilde{S}^{8} \left( - \frac{3274704127}{28}
               + \frac{7211558817317}{25200} \tilde{X}
               - \frac{55615265929}{225} \tilde{X}^{2}
               + \frac{402647627639}{4200} \tilde{X}^{3}  \right. \nonumber \\
    &  &   \left.  \, \hspace{0.6cm} \, - \frac{761509408}{45} \tilde{X}^{4}
               + \frac{692470907}{630} \tilde{X}^{5}  \right)  \nonumber  \\
    & + & \tilde{S}^{10} \left( - \frac{4342880411}{450}
               + \frac{12820511189}{1050} \tilde{X}
               - \frac{1055741186}{225} \tilde{X}^{2}
               + \frac{3487385299}{6300} \tilde{X}^{3}   \right)  \nonumber \\
    & + & \tilde{S}^{12} \left( - \frac{89079832}{675}
               + \frac{1895391191}{37800} \tilde{X}       \right) \, ,  \\
  \nonumber  \\
Z_{6} & = &  \frac{2310525}{8} \tilde{X} - \frac{4808540583}{100} \tilde{X}^{2}
     + \frac{252517854951}{400} \tilde{X}^{3} 
     - \frac{11473454766573}{4900} \tilde{X}^{4}  \nonumber  \\
    &  & + \frac{143434835467311}{39200} \tilde{X}^{5}
     - \frac{429688246765}{147} \tilde{X}^{6}
     + \frac{9794517932561}{7350} \tilde{X}^{7}
     - \frac{72661274793}{196} \tilde{X}^{8}   \nonumber  \\
    &  & + \frac{2312186142587}{35280} \tilde{X}^{9}
     - \frac{11845630792}{1575} \tilde{X}^{10}
     + \frac{2067628712}{3675} \tilde{X}^{11}
     - \frac{127687796}{4725} \tilde{X}^{12}   \nonumber  \\
    &  & + \frac{105557789}{132300} \tilde{X}^{13}
     - \frac{48128}{3675} \tilde{X}^{14}
     + \frac{3008}{33075} \tilde{X}^{15}       \nonumber  \\
    & + & \tilde{S}^{2} \left( - \frac{4808540583}{200}
            + \frac{252517854951}{200} \tilde{X}
            - \frac{126208002432303}{9800} \tilde{X}^{2}
            + \frac{1864652861075043}{39200} \tilde{X}^{3} \right. \nonumber \\
      &  &  \, \hspace{0.6cm} \, - \frac{8164076688535}{98} \tilde{X}^{4}
            + \frac{19589035865122}{245} \tilde{X}^{5}
            - \frac{17947334873871}{392} \tilde{X}^{6}
            + \frac{580358721789337}{35280} \tilde{X}^{7}  \nonumber  \\
      &  &  \, \hspace{0.6cm} \, - \frac{5999811996148}{1575} \tilde{X}^{8}
            + \frac{2104846028816}{3675} \tilde{X}^{9}
            - \frac{86891545178}{1575} \tilde{X}^{10}
            + \frac{143875266407}{44100} \tilde{X}^{11}  \nonumber  \\
      &  &  \left. \, \hspace{0.6cm} \, - \frac{393903616}{3675} \tilde{X}^{12}
            + \frac{49259008}{33075} \tilde{X}^{13}       \right)  \nonumber \\
    & + & \tilde{S}^{4} \left( - \frac{11473454766573}{4900}
               + \frac{2438392202944287}{78400} \tilde{X}
               - \frac{6445323701475}{49} \tilde{X}^{2}
              + \frac{313424573841952}{1225} \tilde{X}^{3} \right. \nonumber \\
      &  &   \, \hspace{0.6cm} \, - \frac{26376042749859}{98} \tilde{X}^{4}
               + \frac{3928404256255313}{23520} \tilde{X}^{5}
               - \frac{100972156871008}{1575} \tilde{X}^{6}
               + \frac{57062417193776}{3675} \tilde{X}^{7}  \nonumber \\
      &  & \left.  \, \hspace{0.6cm} \, - \frac{148756282340}{63} \tilde{X}^{8}
               + \frac{3832908876379}{17640} \tilde{X}^{9}
               - \frac{13467610112}{1225} \tilde{X}^{10}
               + \frac{2574414848}{11025} \tilde{X}^{11}   \right) \nonumber \\
    & + & \tilde{S}^{6} \left( - \frac{7304700195005}{588}
               + \frac{303630055909391}{3675} \tilde{X}
               - \frac{9736610822262}{49} \tilde{X}^{2}
            + \frac{2078655342185713}{8820} \tilde{X}^{3}  \right. \nonumber \\
      &  &  \, \hspace{0.6cm} \, - \frac{245859028495658}{1575} \tilde{X}^{4}
               + \frac{224530004722216}{3675} \tilde{X}^{5}
               - \frac{13559550120628}{945} \tilde{X}^{6}
               + \frac{13054120650052}{6615} \tilde{X}^{7}  \nonumber  \\
      &  & \left.  \, \hspace{0.6cm} \, - \frac{107026432768}{735} \tilde{X}^{8}
               + \frac{29281785088}{6615} \tilde{X}^{9}  \right)  \nonumber  \\
    & + & \tilde{S}^{8} \left( - \frac{2252499518583}{196}
               + \frac{1597720624527617}{35280} \tilde{X}
               - \frac{104821986878408}{1575} \tilde{X}^{2}
             + \frac{177310534011916}{3675} \tilde{X}^{3}  \right. \nonumber \\
      &  &  \left.  \, \hspace{0.6cm} \, - \frac{5960977068464}{315} \tilde{X}^{4}
               + \frac{35848797395657}{8820} \tilde{X}^{5}
               - \frac{328499926016}{735} \tilde{X}^{6}
               + \frac{129412835072}{6615} \tilde{X}^{7}  \right)  \nonumber \\
    & + & \tilde{S}^{10} \left( - \frac{4092665438636}{1575}
               + \frac{22582640792464}{3675} \tilde{X}
               - \frac{8264177610763}{1575} \tilde{X}^{2}
            + \frac{180539814396049}{88200} \tilde{X}^{3}  \right. \nonumber \\
      &  &  \left. \, \hspace{0.6cm} \, - \frac{451257055744}{1225} \tilde{X}^{4}
             + \frac{271877622784}{11025} \tilde{X}^{5}  \right)  \nonumber \\
    & + & \tilde{S}^{12} \left( - \frac{697303053956}{4725}
               + \frac{98123248362941}{529200} \tilde{X}
               - \frac{263501425664}{3675} \tilde{X}^{2}
             + \frac{284503269632}{33075} \tilde{X}^{3} \right)  \nonumber  \\
    & + & \tilde{S}^{14} \left( - \frac{5592287104}{3675}
               + \frac{19264982656}{33075} \tilde{X}     \right)  \, ,
\end{eqnarray}
where $\tilde{X}$ and $\tilde{S}$ are defined by equations (12) and (13), respectively.  In equation (15), the first term corresponds to the first-odrder contribution and the second term corresponds to the second-order contribution of the multiple scattering.  In deriving equation (15), we have used the following identity relation for $y$:
\begin{eqnarray}
\sigma_{T} \, \int_{0}^{\ell} d \ell_{1} N_{e}(\ell_{1})  \, \sigma_{T} \, \int_{0}^{\ell_{1}} d \ell_{2} N_{e}(\ell_{2}) & = & \frac{1}{2} \left( \sigma_{T} \, \int_{0}^{\ell} d \ell_{1} N_{e}(\ell_{1}) \right)^{2}  \,   \nonumber  \\
& = & \frac{1}{2} \, y^{2}  \, . 
\end{eqnarray}
We have also neglected terms higher than $O(\theta_{e}^6)$ in the $y^{2}$ contributions in equation (15).  It is important to note that equation (15) satisfies the photon number conservation.  

  We note that the lowest-order term $Z_{0}$ has been already obtained by Fabbri (1981) and also by Sazonov \& Sunyaev (1998).  The present calculation has produced the relativistic correction terms $Z_{1}$, $Z_{2}$, $Z_{3}$, $Z_{4}$, $Z_{5}$, and $Z_{6}$.

  We will also calculate the multiple scattering effect to the thermal Sunyaev-Zeldovich effect by numerically integrating the collision term of the Boltzmann equation.  Here we assume that the radiation field after the first Compton scattering is isotropic.  The original Boltzmann equation for the photon distribution function $n(\omega)$ is given by (Itoh et al. 1998)
\begin{equation}
\frac{\partial n(\omega)}{\partial t} =  -2 \int \frac{d^{3}p}{(2\pi)^{3}} d^{3}p^{\prime} d^{3}k^{\prime} \, W \, f(E) \,
\left\{ [1 + n(\omega^{\prime})] n(\omega) -  [1 + n(\omega)] n(\omega^{\prime}) e^{ \Delta x }  \right\} \, .
\end{equation}
By inserting equation (14) in equation (24) we obtain
\begin{eqnarray}
\frac{\partial n(\omega)}{\partial t} & = & -2 \int \frac{d^{3}p}{(2\pi)^{3}} d^{3}p^{\prime} d^{3}k^{\prime} \, W \, f(E) \,
\left\{ [1 + n_{0}(\omega^{\prime})] n_{0}(\omega) -  [1 + n_{0}(\omega)] n_{0}(\omega^{\prime}) e^{ \Delta x }  \right\} \, \nonumber  \\
& & -2 \int \frac{d^{3}p}{(2\pi)^{3}} d^{3}p^{\prime} d^{3}k^{\prime} \, W \, f(E) \,
\left\{ n_{0}(\omega^{\prime}) \frac{ \Delta n(\omega^{\prime})}{n_{0}(\omega^{\prime})} \left[ n_{0}(\omega) \left(1 - e^{ \Delta x} \right) - e^{ \Delta x } \right]  \right. \nonumber \\
& & \hspace{4.5cm} \left. + n_{0}(\omega) \frac{\Delta n(\omega)}{n_{0}(\omega)} \left[ 1 + n_{0}(\omega^{\prime}) \left(1 - e^{ \Delta x} \right) \right]  \right\}  \, .
\end{eqnarray}
The first term corresponds to the single scattering contribution which has been calculated by Itoh et al. (1998).  The second term corresponds to the double scattering contribution.  For $\Delta n(\omega^{\prime})/n_{0}(\omega^{\prime})$ we use the numerical results obtained by Itoh et al. (1998) and calculate the double scattering contribution by numerical integration of the collision term of the Boltzmann equation.

  With equation (15), we define the distortion of the spectral intensity as follows:
\begin{eqnarray}
\Delta I & = & \frac{X^3}{e^{X} - 1} \, \frac{\Delta n(X)}{n_{0}(X)}  \, = \, \Delta I_{1} \, + \, \Delta I_{2} \, .
\end{eqnarray}
The first term $\Delta I_{1}$ contains a factor $y$, whereas the second term $\Delta I_{2}$ contains a factor $y^{2}$.  In Figure 1 we show $\Delta I_{2}/y^{2}$ as a function of $X$ for the case $k_{B} T_{e}$ = 5keV.  For $k_{B} T_{e} \leq$ 5keV it is found that results of the Fokker-Planck expansion approximation perfectly agree with that obtained by the numerical integration of the collision term of the Boltzmann equation for the entire region of $X \leq 20$.  For $k_{B} T_{e} \leq$ 5keV the convergence of the series with respect to the relativistic temperature parameter $\theta_{e}$ is relatively fast.  Therefore it is sufficient to include up to $\theta_{e}^{4} Z_{4}$ terms in equation (15).  In Figure 2 we show $\Delta I_{2}/y^{2}$ for the case $k_{B} T_{e}$ = 10keV.  For this temperature region the convergence of the series expansion is slow for large values of $X$.  The Fokker-Planck expansion aprroximation is valid for $X \leq 10$.  For higher temperature region, the convergence is even worse.  For $k_{B} T_{e}$ = 15keV the expansion approximation is valid for the region of $X \leq 4$.

  In order to estimate the relative importance of the multiple scattering contribution, we now define the following ratio:
\begin{eqnarray}
\Gamma & \equiv & \frac{\Delta I_{2}/y^{2}}{\Delta I_{1}/y}   \, .  
\end{eqnarray}
In Figure 3 we show $\Gamma$ for $X=5$ as a function of the electron temperature $T_{e}$.  It is clear that $\Gamma$ increases with a negative sign as the temperature of the cluster of galaxies increases.  The numerical result (solid curve) shows that the maximum contribution is $\Gamma \approx -0.2$ at $k_{B} T_{e}=15$keV.  However, the multiple scattering contribution is small because of a further factor $y$.  Namely, for the cluster of galaxies of $k_{B} T_{e}=15$keV, we have
\begin{eqnarray}
\frac{\Delta I_{2}}{\Delta I_{1}} & = & y \, \Gamma \, \approx -0.2 \, y  \, \approx  -0.2 \% \,  ,
\end{eqnarray}
where we used a typical value $y \approx 0.01$ of the galaxy clusters.  Therefore the maximum effect of the multiple scattering contribution is $-0.2\%$ of the single scattering contribution for the observed high-temperature galaxy clusters.

  In the Rayleigh--Jeans limit where $X \rightarrow 0$, equation (15) is further simplified:
\begin{eqnarray}
\frac{\Delta n(X)}{n_{0}(X)} & = & - 2 \, y \, \theta_{e} \, \left( 1 - \frac{17}{10} \theta_{e} + \frac{123}{40} \theta_{e}^{2} - \frac{1989}{280} \theta_{e}^{3}
 + \frac{14403}{640} \theta_{e}^{4} \, - \frac{20157}{224} \theta_{e}^{5} \,  +  \, \frac{423951}{1024} \theta_{e}^{6} \, \right)  \nonumber  \\
&  &  + 2 \, y^2 \, \theta_{e}^2 \, \left( 1 - \frac{17}{5} \theta_{e} + \frac{226}{25} \theta_{e}^{2} - \frac{34527}{1400} \theta_{e}^3 +
 \frac{13758}{175} \theta_{e}^4 \, - \, \frac{1344789}{4480} \theta_{e}^{5} \,
 + \, \frac{25927827}{19600} \theta_{e}^{6} \, \right)  \, .
\end{eqnarray}
With equation (29) we have the multiple scattering contribution for $k_{B}T_{e}=$15keV and $y=0.01$ as follows:
\begin{eqnarray}
\frac{\Delta I_{2}}{\Delta I_{1}} & \approx & -y \, \theta_{e} \, \approx  -0.03 \% \,  .
\end{eqnarray}
In the Rayleigh--Jeans region the multiple scattering contribution is safely neglected.

\section{Discussions and concluding remarks}

  We have calculated the relativistic corrections to the multiple scattering contribution to the Sunyaev-Zel'dovich effect in the isotropic approximation by extending the formalism developed in our previous papers as well as by Fabbri (1981) and Sazonov \& Sunyaev (1998).  We have also calculated the multiple scattering effect to the Sunyaev-Zeldovich effect by numerically integrating the collision term of the Boltzmann equation.  Our approach is complementary to the Monte Carlo calculation of Molnar \& Birkinshaw (1999).  We have estimated the accuracy of the Fokker-Planck expansion approximation by comparing with the result obtained by the numerical integration of the collision term of the Boltzmann equation.  We have found that Fokker-Planck expansion approximation is valid for the enire region of $X \leq 20$ for $k_{B} T_{e} \leq$ 5keV.  However, for $k_{B} T_{e} \leq$ 10keV, valid region is limited to $X \leq 10$.  For higher temperature region, the convergence is even worse.  For $k_{B} T_{e}$ = 15keV the expansion approximation is valid for the region of $X \leq 4$.

  From the results presented in the previous section it is clear that the multiple scattering contribution $\Delta I_{2}$ is very small compared with the single scattering contribution $\Delta I_{1}$.  For high-temperature galaxy clusters of $k_{B} T_{e} \approx 15$keV, we obtain the ratio $\Delta I_{2}/\Delta I_{1} \approx -0.2\%$ at $X=5$.  In the Rayleigh--Jeans region we have $\Delta I_{2}/\Delta I_{1} \approx -0.03\%$.  Therefore it is concluded that the multiple scattering contribution to the thermal Sunyaev-Zel'dovich effect for galaxy clusters can be safely neglected.  The reader is therefore referred to our previous four papers which deal with the single scattering contribution in detail.

\section*{Acknowledgments}

  We wish to thank Dr. S. Y. Sazonov and Dr. S. M. Molnar for valuable communication.  This work is financially supported in part by the Grant-in-Aid of Japanese Ministry of Education, Science, Sports, and Culture under the contract \#10640289.

\newpage

\bigskip

\noindent
{\bf \large FIGURE CAPTIONS}

\begin{itemize}

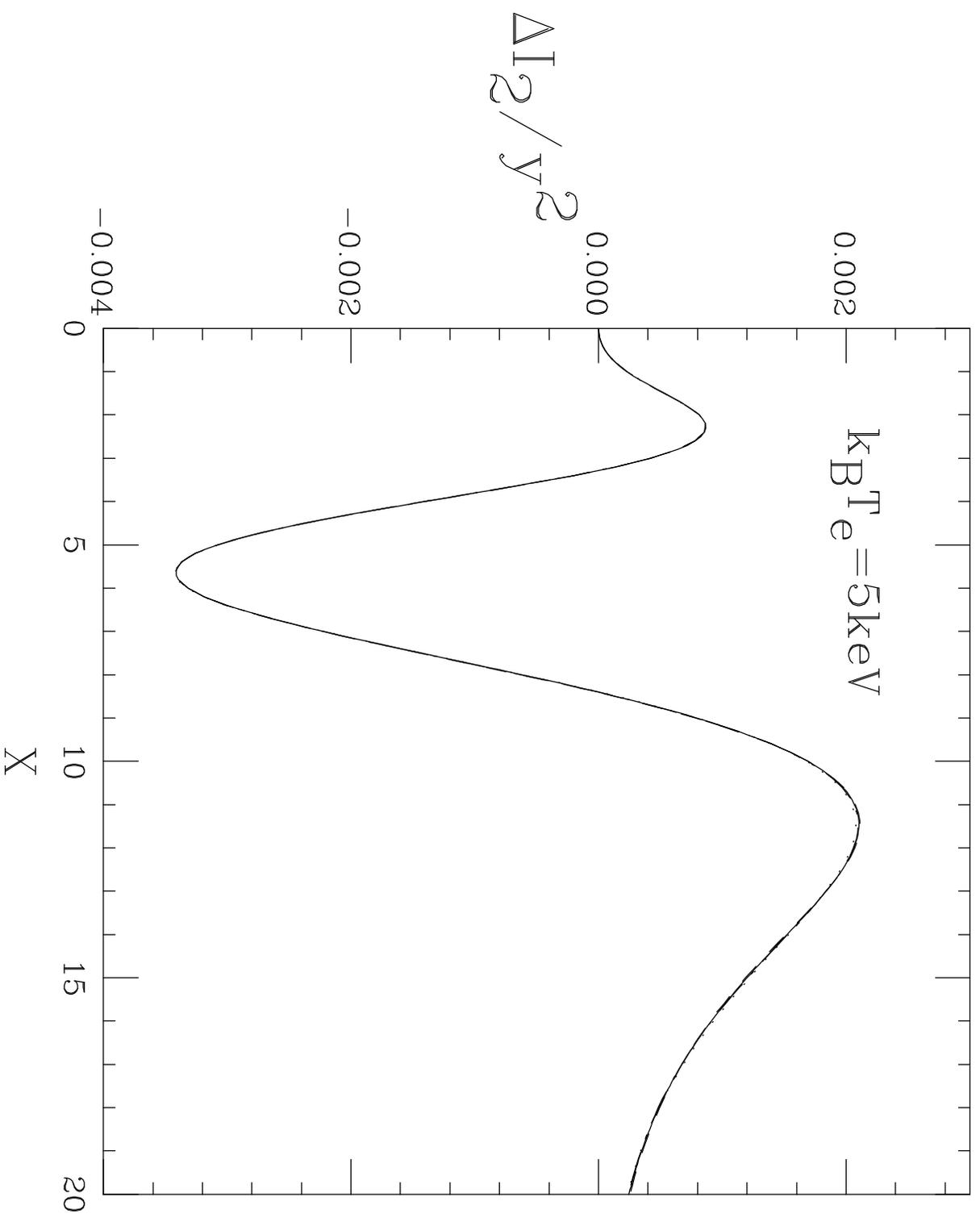
\item Figure 1. The spectral intensity distortion $\Delta I_{2}/y^{2}$ as a function of $X$ for the case $k_{B} T_{e}$ = 5keV.  The dotted curve shows the contribution up to $\theta_{e}^{4} Z_{4}$.  The dashed curve shows the contribution up to $\theta_{e}^{5} Z_{5}$.  The dash-dotted curve shows the full contribution up to $\theta_{e}^{6} Z_{6}$.  The solid curve shows the results of the numerical integration.

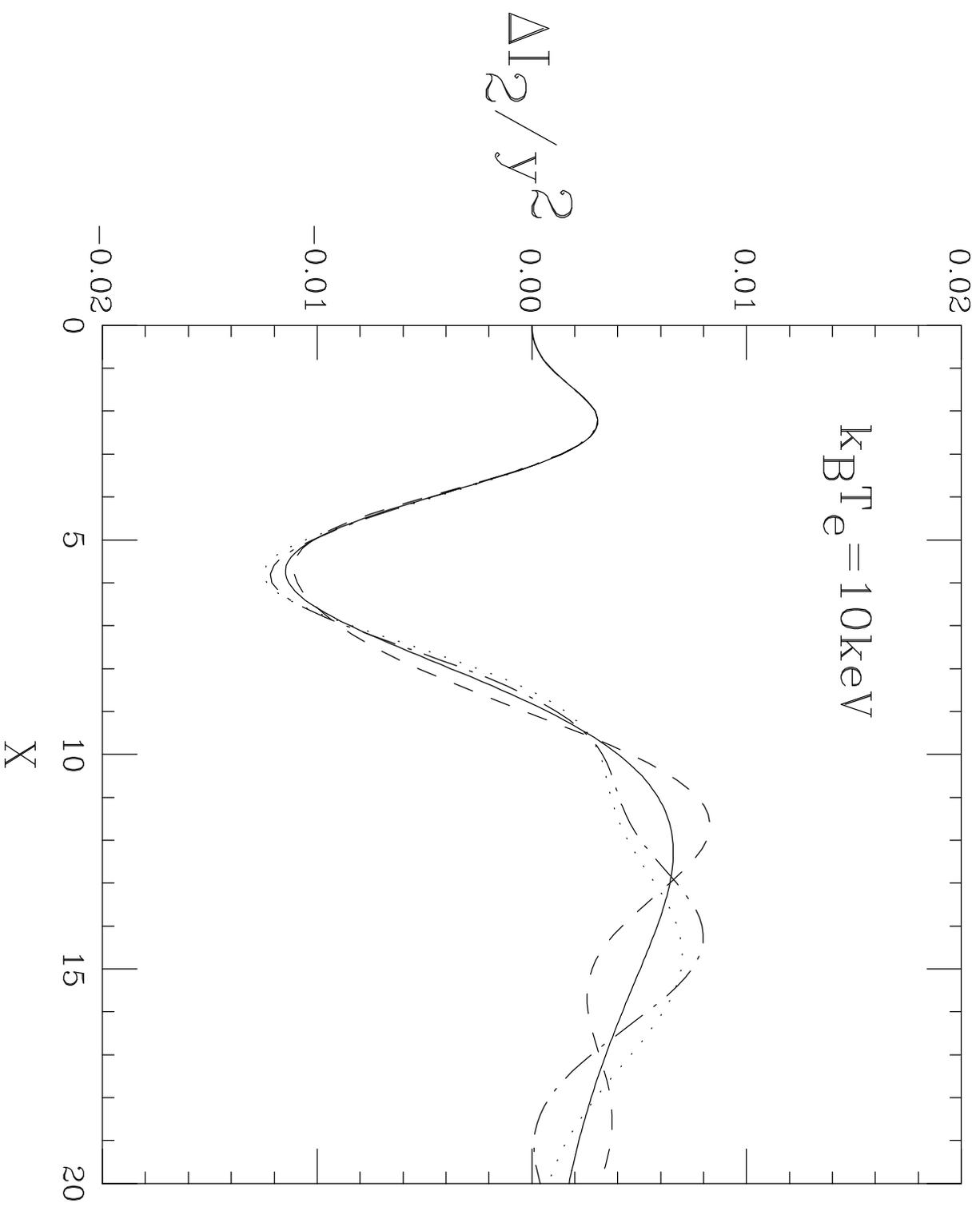
\item Figure 2. Same as Figure 1 except for $k_{B} T_{e}$ = 10keV.  The dotted curve shows the contribution up to $\theta_{e}^{4} Z_{4}$.  The dashed curve shows the contribution up to $\theta_{e}^{5} Z_{5}$.  The dash-dotted curve shows the full contribution up to $\theta_{e}^{6} Z_{6}$.  The solid curve shows the results of the numerical integration.

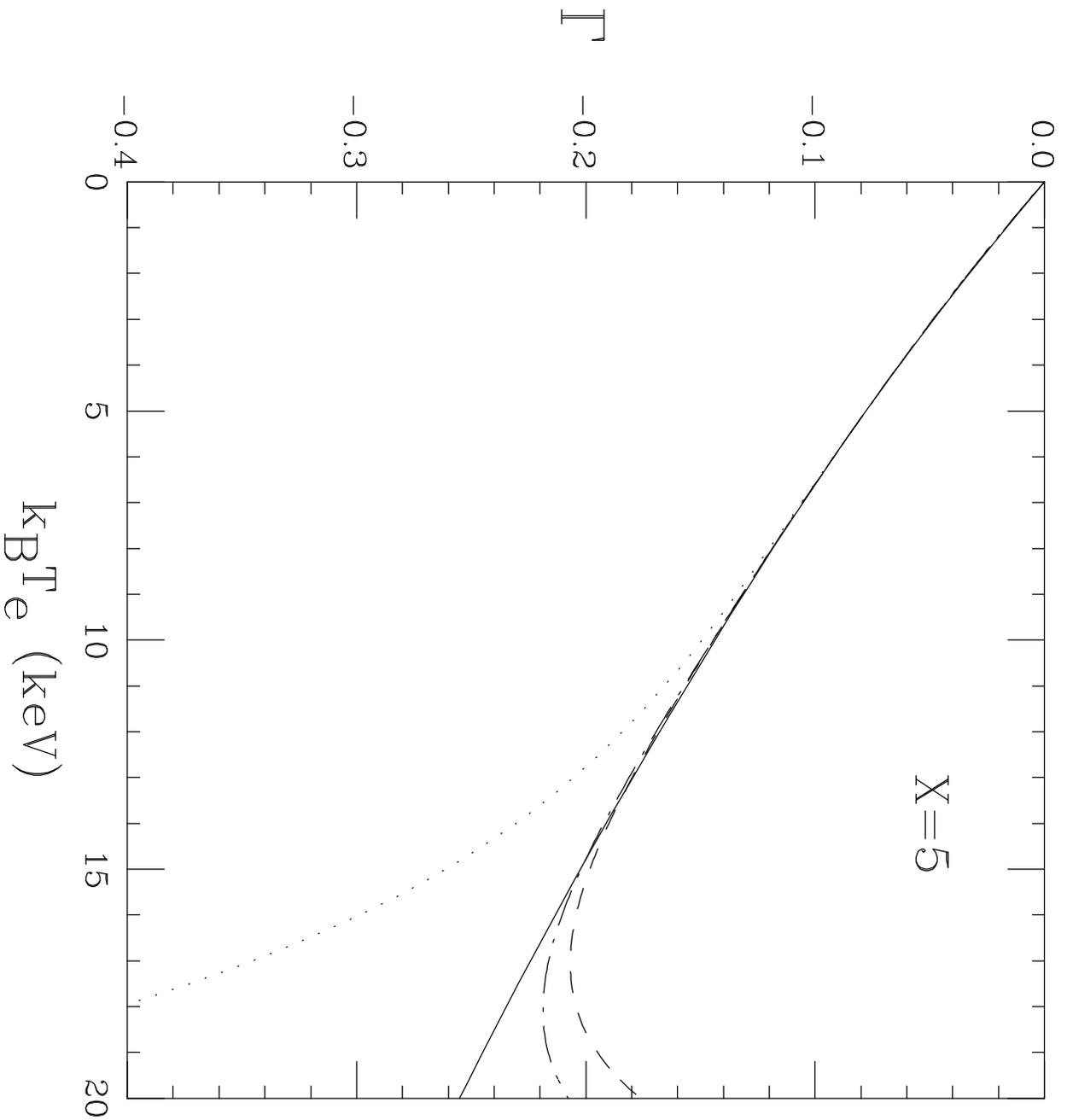
\item Figure 3. The ratio $\Gamma$ as a function of $k_{B} T_{e}$ for a fixed value of $X=5$.  The dotted curve shows the contribution up to $\theta_{e}^{4} Z_{4}$.  The dashed curve shows the contribution up to $\theta_{e}^{5} Z_{5}$.  The dash-dotted curve shows the full contribution up to $\theta_{e}^{6} Z_{6}$.  The solid curve shows the results of the numerical integration.
\end{itemize}

\bigskip

\bsp

\label{lastpage}


\begin{thebibliography}{99}
\bibitem{Arna94} Arnaud K. A., Mushotzky R. F., Ezawa H., Fukazawa Y., Ohashi T., Bautz M. W., Crewe G. B., Gendreau K. C., Yamashita K., Kamata Y., Akimoto F., 1994, ApJ, 436, L67
\bibitem{BLP82} Berestetskii V. B., Lifshitz E. M., Pitaevskii L. P., 1982, { \it Quantum Electrodynamics} (Oxford: Pergamon)
\bibitem{Birk79} Birkinshaw M., 1979, MNRAS, 187, 847
\bibitem{Birk99} Birkinshaw M., 1999, Physics Reports, 310, 97
\bibitem{BH94} Birkinshaw M., Hughes J. P., 1994, ApJ, 420, 33
\bibitem{BHA91} Birkinshaw M., Hughes J. P., Arnaud K. A., 1991, ApJ, 379, 466
\bibitem{BY76} Buchler J. R., Yueh W. R., 1976, ApJ, 210, 440
\bibitem{CDZ79} Cavaliere A., Danese L., De Zotti G., 1979, A\&A, 75, 322
\bibitem{CL98} Challinor A., Lasenby A., 1998, ApJ, 499, 1
\bibitem{Davi93} David L. P., Slyz A., Jones C., Forman W.,  Vrtilek S. D., 1993, ApJ, 412, 479
\bibitem{Fabb81} Fabbri R., 1981, Ap. Space Sci., 77, 529
\bibitem{Furu97} Furuzawa A., Tawara Y., Kunieda H., Yamashita K., Sonobe T., Tanaka Y., Mushotzky R., 1998, ApJ, 504, 35
\bibitem{Gunn78} Gunn J. E., 1978, in Observational Cosmology, 1, ed. A. Maeder, L. Martinet \& G. Tammann (Sauverny: Geneva Obs.)
\bibitem{HLRG95} Herbig T., Lawrence C. R., Readhead A. C. S., Gulkus S., 1995, ApJ, 449, L5
\bibitem{Holz97} Holzapfel W. L. et al., 1997, ApJ, 480, 449
\bibitem{HB98} Hughes J. P., Birkinshaw M., 1998, ApJ, 501, 1
\bibitem{IKN98} Itoh N., Kohyama Y., Nozawa S., 1998, ApJ, 502, 7
\bibitem{INK00} Itoh N., Nozawa S., Kohyama Y., 2000, ApJ, 533, 588
\bibitem{Jone95} Jones M., 1995, Astrophys. Lett. Commun., 6, 347
\bibitem{Koma99} Komatsu E., Kitayama T., Suto Y., Hattori M., Kawabe R., Matsuo H., Schindler S., Yoshikawa K., 1999, ApJ, 516, L1
\bibitem{Komp57} Kompaneets A. S., 1957, Soviet Physics JETP, 4, 730
\bibitem{LML} Loeb A., McKee C. F., Lahav O., 1991, ApJ, 374, 44
\bibitem{Mark98} Markevitch M., 1998, ApJ, 504, 27
\bibitem{Mark96} Markevitch M., Mushotzky R., Inoue H., Yamashita K., Furuzawa A., Tawara Y., 1996, ApJ, 456, 437
\bibitem{Mark94} Markevitch M., Yamashita K., Furuzawa A., Tawara Y., 1994, ApJ, 436, L71
\bibitem{MB99} Molnar S. M., Birkinshaw M., 1999, ApJ, 523, 78
\bibitem{MBRH95} Myers S. T., Baker J. E., Readhead A. C. S., Herbig T., 1995, preprint
\bibitem{MS97} Mushotzky R. F., Scharf C. A., 1997, ApJ, 482, L13
\bibitem{NIKK00} Nozawa S., Itoh N., Kawana Y., Kohyama Y., 2000, ApJ, 536, 31
\bibitem{NIK98} Nozawa S., Itoh N., Kohyama Y., 1998, ApJ, 508, 17
\bibitem{Rees00} Reese E. D. et al., 2000, ApJ, 533, 38
\bibitem{Reph95} Rephaeli Y., 1995, ApJ, 445, 33
\bibitem{RY97} Rephaeli Y., Yankovitch, D., 1997, ApJ, 481, L55
\bibitem{SS98} Sazonov S. Y., Sunyaev R. A., 1998, Astronomy Letters 24, 553
\bibitem{SW78} Silk J. I., White S. D. M., 1978, ApJ, 226, L103
\bibitem{Steb97} Stebbins A., 1997, preprint astro-ph/9705178
\bibitem{SZ72} Sunyaev R. A., Zel'dovich Ya. B., 1972, Comments Astrophys. Space Sci., 4, 173
\bibitem{SZ80a} Sunyaev R. A., Zel'dovich Ya. B., 1980a, ARA\&A, 18, 537
\bibitem{SZ80b} Sunyaev R. A., Zel'dovich Ya. B., 1980b, MNRAS, 190, 413
\bibitem{SZ81} Sunyaev R. A., Zel'dovich Ya. B., 1981, Astrophysics and Space Physics Reviews, 1, 1
\bibitem{Tuck98} Tucker W., Blanco P., Rappoport S., David L., Fabricant, D., Falco, E. E., Forman, W., Dressler, A., \& Ramella, M. 1998, ApJ, 496, L5
\bibitem{Weym65} Weymann R., 1965, Phys. Fluid, 8, 2112
\bibitem{Wrig79} Wright E. L., 1979, ApJ, 232, 348
\bibitem{ZS69} Zel'dovich Ya. B., Sunyaev R. A., 1969, Astrophys. Space Sci., 4, 301

\end{thebibliography}
\end{document}